\documentclass[12pt,prl,english,twocolumn,reprint]{revtex4-1}
\usepackage{times}
\usepackage{array}
\usepackage{longtable}
\usepackage{float}
\usepackage{amsmath}
\usepackage{graphicx}
\usepackage{amssymb}
\usepackage{color}
\usepackage[FIGTOPCAP]{subfigure}
\usepackage[normalem]{ulem}

\makeatletter
\usepackage{fancyhdr}
\usepackage{babel}
\makeatother
\begin{document}

\title{The constraint of plasma power balance on runaway avoidance}

\author{Christopher J. McDevitt}
\affiliation{Nuclear Engineering Program, University of Florida, Gainesville, FL 32611}
\author{Xian-Zhu Tang, Christopher J. Fontes, Prashant Sharma}

\affiliation{Los Alamos National Laboratory, Los Alamos, NM 87545}
\author{Hyun-Kyung Chung}
\affiliation{Korea Institute of Fusion Energy,  169-148 Gwahak-ro, Yuseong-gu, Daejeon 34133, Korea}

\date{\today}

\begin{abstract}


  In a post-thermal-quench plasma, mitigated or unmitigated, the
  plasma power balance is mostly between collisional or Ohmic heating
  and plasma radiative cooling. In a plasma of atomic mixture
  $\{n_\alpha\}$ with $\alpha$ labeling the atomic species, the power
  balance sets the plasma temperature, ion charge state distribution
  $\{n_\alpha^i\}$ with $i$ the charge number, and through the
  electron temperature $T_e$ and ion charge state distribution
  $\{n_\alpha^i\},$ the parallel electric field $E_\parallel.$ Since
  the threshold electric field for runaway avalanche growth $E_{av}$
  is also set by the atomic mixture, ion charge state distribution and
  its derived quantity, the electron density $n_e,$ the plasma power
  balance between Ohmic heating and radiative cooling imposes a
  stringent constraint on the plasma regime for avoiding and
  minimizing runaways when a fusion-grade tokamak plasma is rapidly
  terminated.
  
 \end{abstract}

\maketitle



The fast termination of a fusion-grade plasma in a tokamak reactor is
prone to Ohmic-to-runaway current
conversion~\cite{Hender-etal-nf-2007}, which is made extraordinarily
efficient by the avalanche mechanism~\cite{Sokolov:1979,
  Jayakumar:1993, Rosenbluth:1997} due to the knock-on collisions
between primary runaways and background free and bound
electrons~\cite{Martin:2017, hesslow2019influence,
  mcdevitt2019avalanche}. Such fast shutdowns could be intentional, for safety upon the detection of an inadvertent sub-system fault, for example, or unplanned, as the result of a tokamak
disruption. Disruptions can have a variety of
causes~\cite{deVries-etal-nf-2011} including such a mundane event as a
falling tungsten flake into the plasma. For the relativistic energies
characteristic of runaway electrons (RE), their local deposition on
the first wall can induce severe surface and sub-surface damage of
plasma facing components.  A straightforward and perhaps ideal
approach to mitigate RE damage is to minimize the runaway population
by avoiding the runaway avalanche altogether. This is the so-called
runaway electron avoidance problem in a tokamak plasma.

The most troublesome feature of a fast shutdown, as in a tokamak
disruption, is the ease for a fusion-grade plasma to rid its thermal
energy in comparison with the plasma current.  The so-called thermal
quench (loss of plasma thermal energy) is often one to two orders of
magnitude (if not more) shorter than the current quench (decay of
plasma current)~\cite{Hender-etal-nf-2007}. In a post-thermal-quench
plasma, mitigated or not, the plasma power balance is mostly between
collisional or Ohmic heating and plasma radiation. This is usually the
case because the post-thermal-quench plasma temperature is clamped by
high-Z impurity radiation to be a very low value, likely in the range
of a few electron volts. Radial transport at such low thermal energies
is relatively slow, even in the presence of a stochastic magnetic
field~\cite{ward1992impurity, breizman2019physics}.
The source of high-Z impurities could be divertor/wall materials that are introduced
into the plasma through intense plasma-wall interaction during the
thermal quench when the bulk of the plasma thermal energy is dumped on
the plasma-facing components. In a mitigated thermal quench, high-Z
impurities, such as neon or argon, are deliberately injected into the
plasma via pellets or gas jets.

In the standard scenario where the thermal quench is fast and the
post-thermal-quench plasma is cold and rich in high-Z impurities, an
Ohmic-to-runaway current conversion is inevitable when a finite RE
seed and large amount of plasma current is present. This results in
the formation of a runaway plateau shortly after the thermal quench.
An interesting discovery, from experiments on both
DIII-D~\cite{paz2021novel} and JET~\cite{reux2021demonstration}, is
that the high-Z impurities can be purged by a massive
deuterium injection in the runaway plateau phase. The
resulting, mostly deuterium plasma can expel the REs via a large-scale MHD event leading to a globally stochastic magnetic field.
Since this RE mitigation scheme does not
  rely on the strict avoidance of REs, it offers the possibility of
  simultaneously satisfying competing requirements such as thermal
  quench and RE mitigation. The details of the underlying MHD
instabilities vary in DIII-D and JET
experiments~\cite{bandaru2021magnetohydrodynamic}, but the expectation
that open field lines lead to rapid runaway loss via parallel
streaming is robustly met in both devices. The added benefit is the
experimental observation that the runaways are broadly disbursed onto
the first wall so no appreciable localized heating is detected. The
so-called MHD flush of the runaways after an impurity purge leaves the
possibility that the mostly deuterium plasma could reheat
to sustain an Ohmic current without crossing the avalanche threshold.
This is the topic of the current paper.

\begin{figure}
\begin{centering}
\includegraphics[scale=0.33]{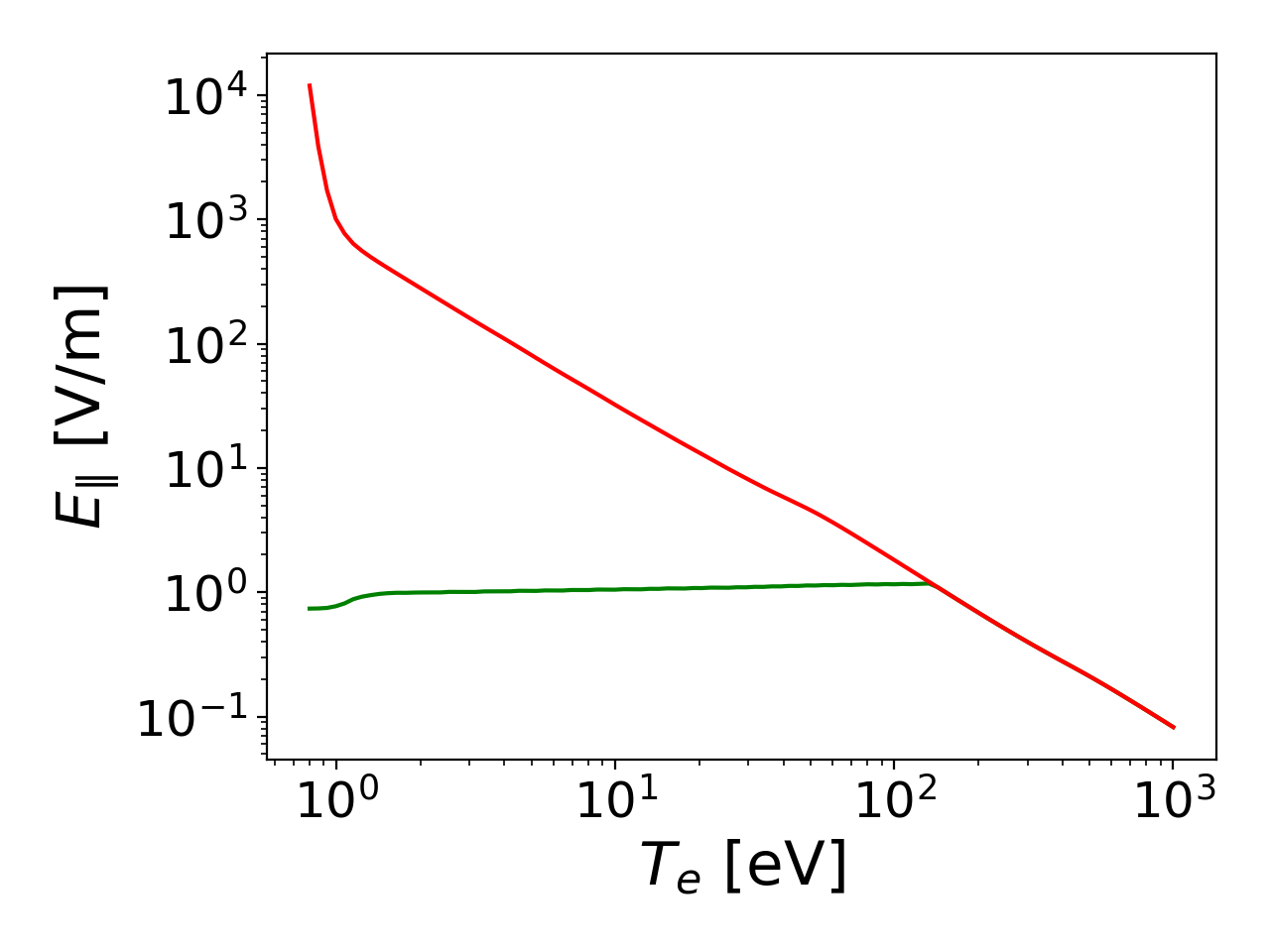}
\par\end{centering}
\caption{Transition between Ohmic and RE roots. The red
  curve indicates the parallel electric field on the Ohmic root, whereas the green curve indicates
  the parallel electric field on the RE root. The temperature at which the curves intersect defines $T_{av}$. The deuterium density was taken to be $n_D=10^{21}$~m$^{-3}$, the neon density $n_{Ne}=10^{19}$~m$^{-3}$, and the current density was taken to be  $j = 2\;\text{MA/m}^2$.}
\label{fig:RE-heating}
\end{figure}

In a plasma of atomic mix $\{n_\alpha\}$ with $\alpha$ labeling the
atomic species, the power balance between Ohmic heating and radiative
cooling sets the plasma temperature, ion charge state distribution
$\{n_\alpha^i\}$ with $i$ the charge number, and through the electron
temperature $T_e$, the ion charge state distribution $\{n_\alpha^i\},$
and the parallel electric field $E_\parallel.$ Since the threshold
electric field for runaway avalanche growth $E_{av}$ is also set by
the atomic mixture, charge state distribution and its derived
quantity, the electron density $n_e,$ the plasma power balance between
Ohmic heating and radiative cooling imposes a stringent constraint on
the plasma regime for avoiding and minimizing runaways when a
fusion-grade tokamak plasma is to be terminated either intentionally
or unintentionally.  Robust RE avoidance can be achieved if Ohmic
heating is able to offset the radiative and transport losses, and
reheat the plasma so the parallel electric field $E_{\parallel}=\eta
j_\Vert$ drops below the runaway avalanche threshold $E_{av}.$ If this
could be maintained over the remainder of the current quench,
effective runaway ``avoidance'' would have been achieved.  The key
question is the critical deuterium density and the fractional neon
impurity density below which such a scenario can be triggered. A
second question is whether the reheated plasma can be placed in the
regime that the Ohmic current quench falls within the known design
constraint for the current quench duration, which in the case of ITER
has an upper bound of 150~milliseconds (ms), for limiting
  the halo current, and a lower bound of 50~ms  in order
  to avoid excessive eddy currents.~\cite{Hender-etal-nf-2007,
  Hollmann-etal-PoP-2015, lehnen2015disruptions}

\begin{figure}
\begin{centering}
  \subfigure[]{\includegraphics[scale=0.24]{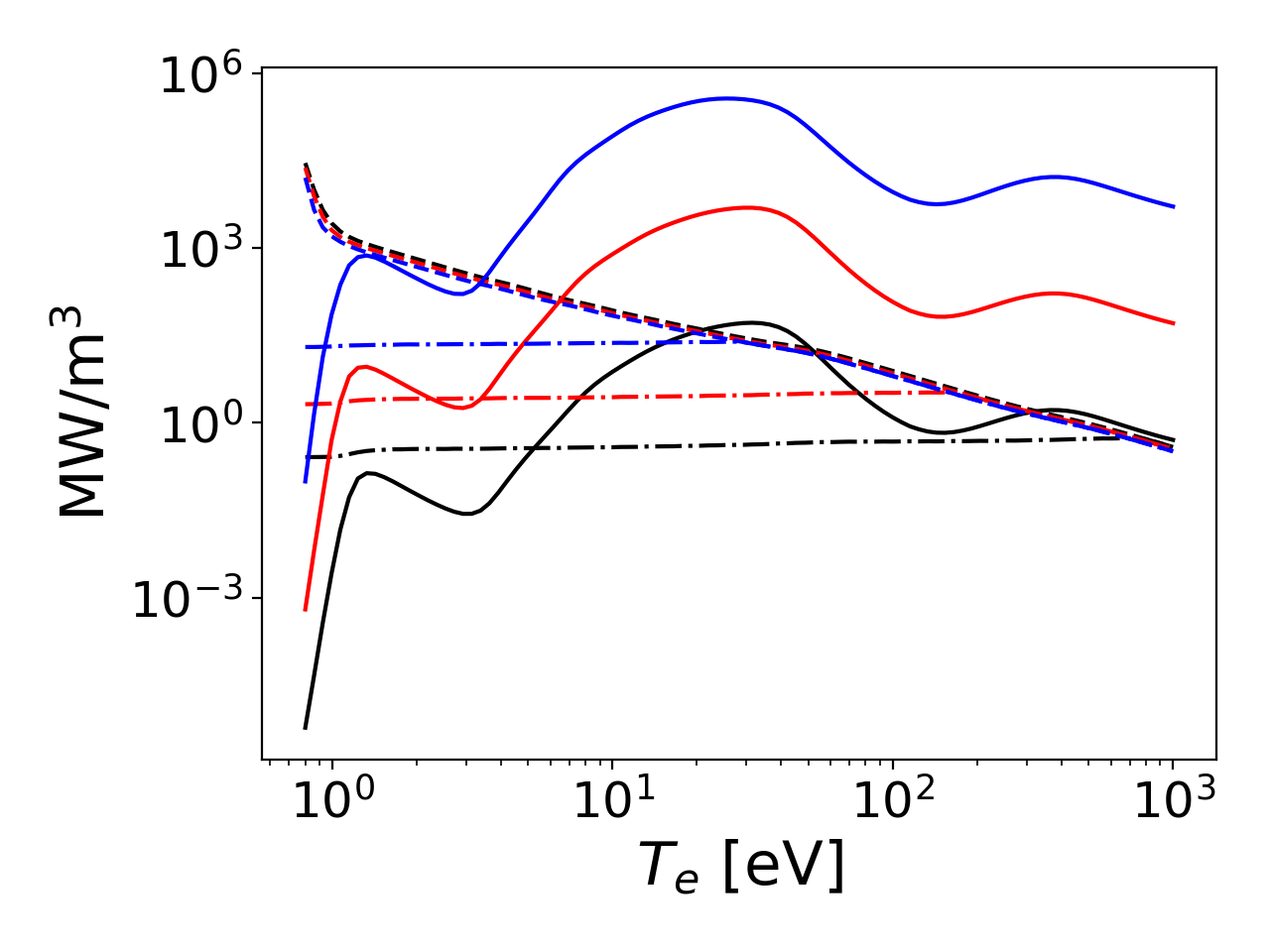}}
  \subfigure[]{\includegraphics[scale=0.24]{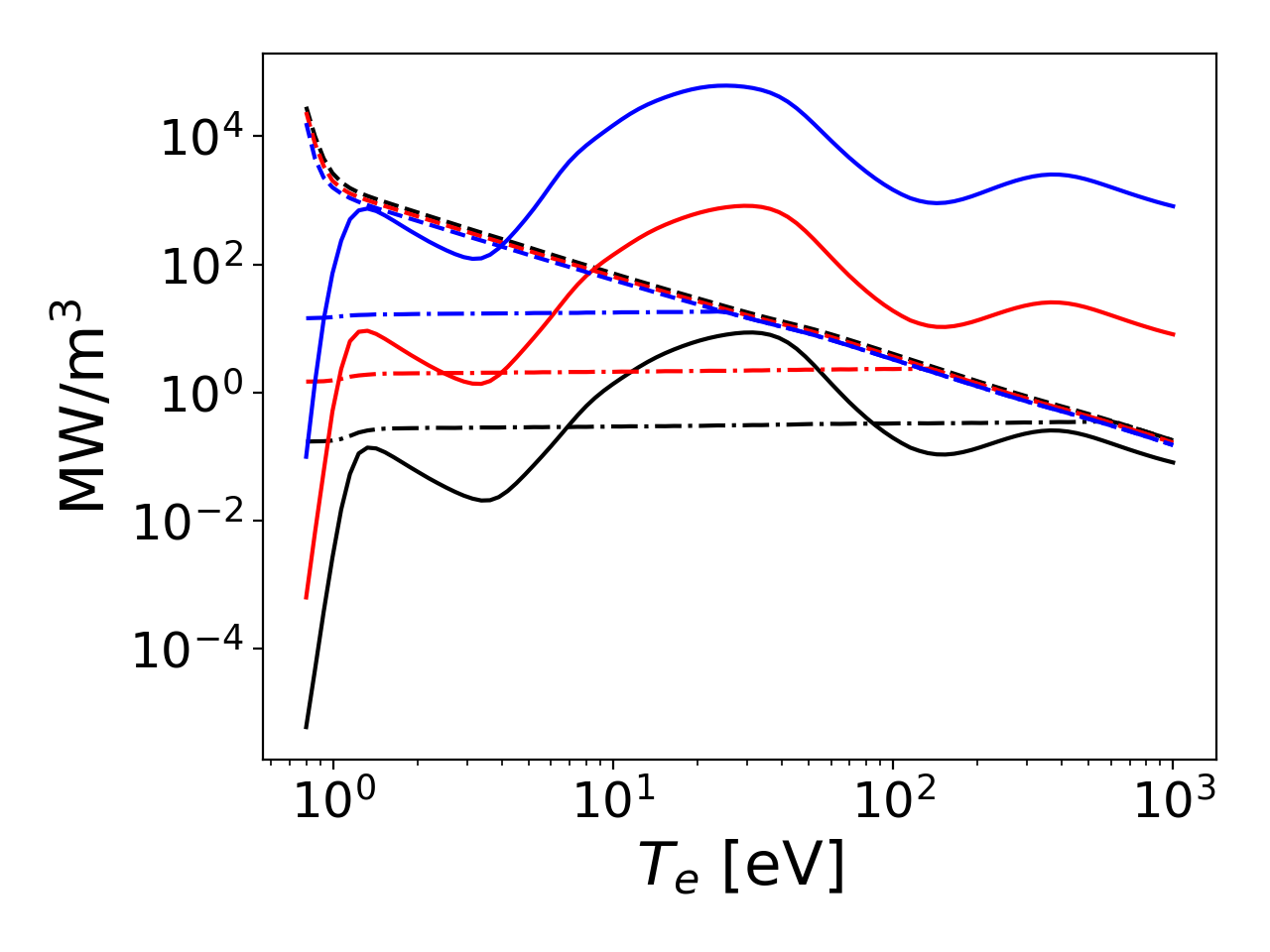}}
  \subfigure[]{\includegraphics[scale=0.24]{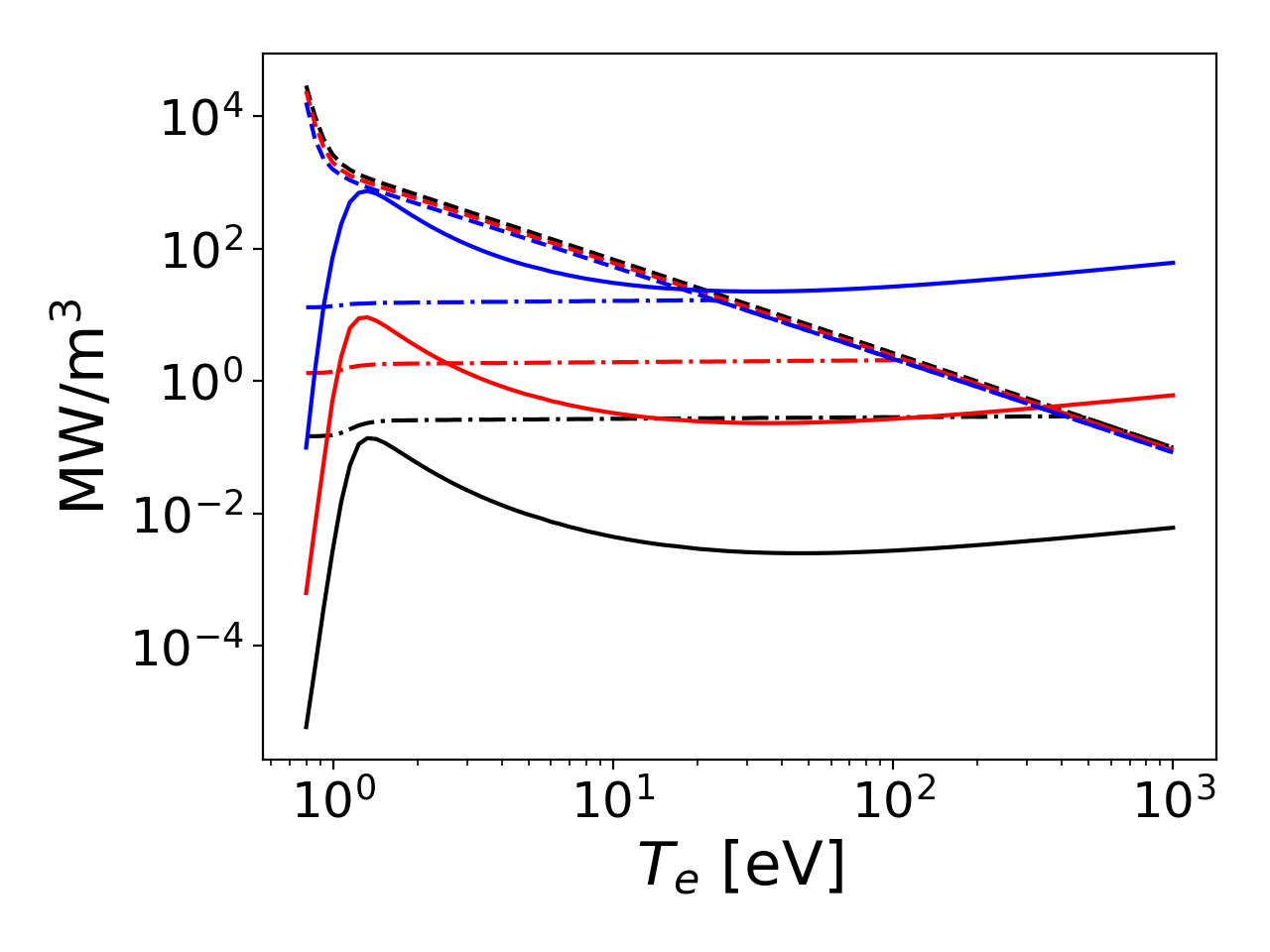}}
  \subfigure[]{\includegraphics[scale=0.24]{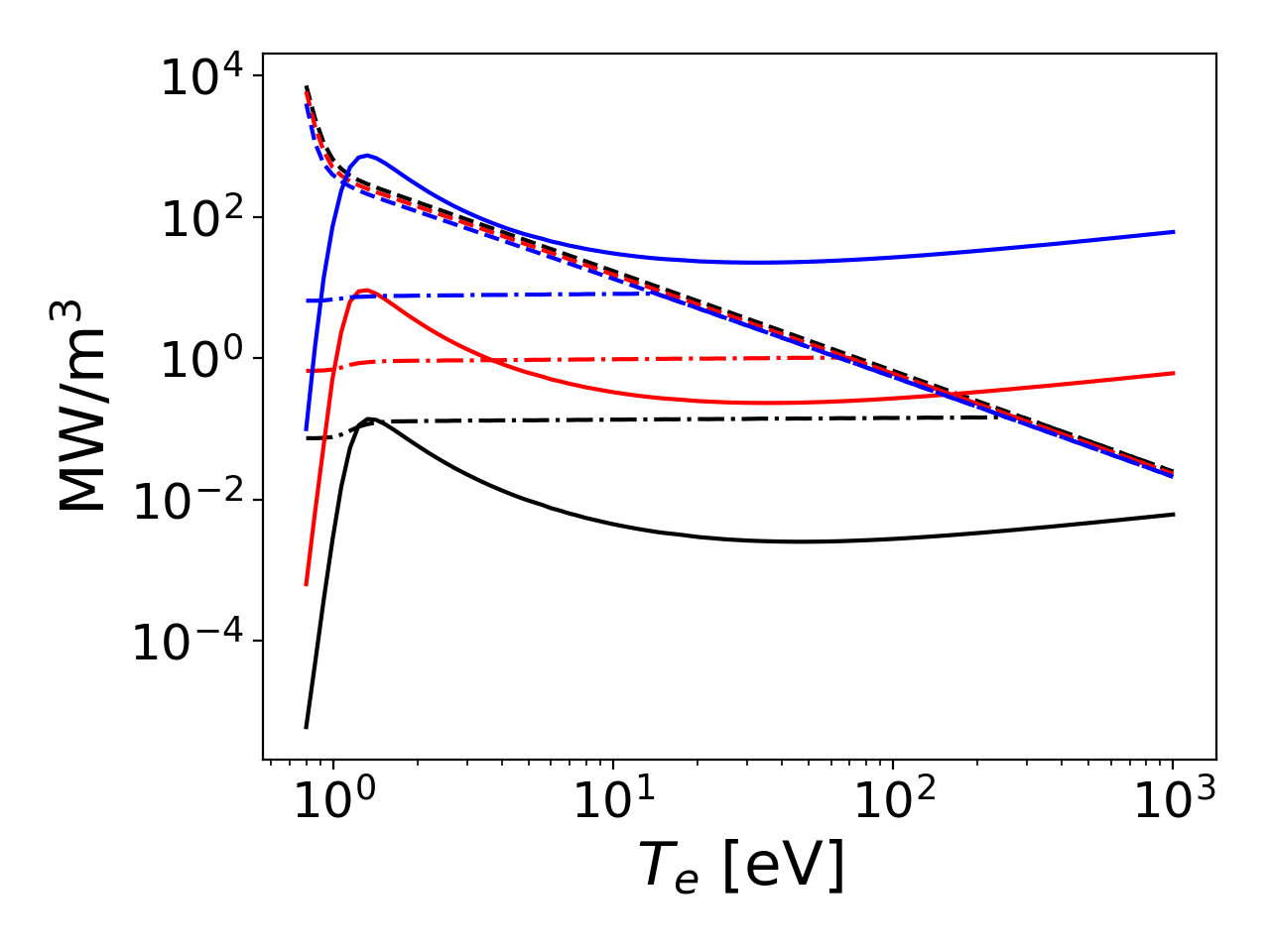}}
\par\end{centering}
\caption{Ohmic heating $\eta j^2$ (dashed lines) with current carried
  by background electrons, collisional heating
  $\mathbf{E}_{av}\cdot\mathbf{j}$ (dashed-dotted lines) with current
  carried by runaway electrons and $E_{av}$ the avalanche threshold
  field, and radiative cooling rate $P_{rad}$ (solid lines) are shown
  as a function of $T_e$ and for three deuterium densities:
  $n_D=10^{20}$~m$^{-3}$ (black), $n_D=10^{21}$~m$^{-3}$ (red), and
  $n_D=10^{22}$~m$^{-3}$ (blue). There are 4 cases shown: (a) $j =
  2$~MA/m$^2, n_{Neon}/n_D = 5$\%; (b) $j = 2$~MA/m$^2, n_{Neon}/n_D = 1$\%;
  (c) $j = 2$~MA/m$^2, n_{Neon}=0;$ (d) $j = 1$~MA/m$^2, n_{Neon}=0.$}
\label{fig:runaway-avoidance}
\end{figure}

This Letter lays out the basic physics considerations underlying the
answers to both questions explained above, which are of practical
importance to a tokamak reactor like ITER. From the plasma power
balance between Ohmic heating and radiative cooling, we find that the
operational space for plasma reheating and runaway avoidance is highly
constrained in terms of the plasma density and the remnant impurity
content. 
This can be illustrated by considering the quasi-steady state parallel
electric field as a function of the electron temperature, an example of
which is plotted in Fig. \ref{fig:RE-heating}. First considering the
case in which a negligible number of runaway electrons are present,
the parallel electric field will be given by $E_\Vert = \eta j_\Vert$,
with $\eta$ the plasma resistivity and $j_\parallel$ the plasma
parallel current density~\cite{comment:force-free}. Noting that the
plasma resistivity scales as $\eta \propto 1/T^{3/2}_e$, the electric
field will decrease rapidly as $T_e$ is increased for a given plasma
current density $j_\Vert$. Once the magnitude of the electric field
has dropped below $E_{av}$, runaway electron amplification by the
avalanche mechanism will no longer be possible. The electron
temperature at which this occurs will be referred to as $T_{av}$. For
temperatures below $T_{av}$, two distinct roots of the system are
present. This can be motivated by considering an Ohm's law, modified to
account for the presence of runaway electrons, of the form:
\[
E_\Vert = \eta \left( j_\Vert - j_{RE}\right)
.
\]
For $j_{RE} \ll j_\Vert$, the electric field can again be approximated
by $E_\Vert \approx \eta j_\Vert$, which yields the red curve shown in
Fig. \ref{fig:RE-heating}. For $T_e < T_{av}$ this root can, however,
be recognized to be unstable. In particular, since $E_\Vert > E_{av}$
when $T_e < T_{av}$, any seed RE population present in the plasma will
be amplified by the avalanche mechanism. As a larger fraction of the
plasma current is carried by REs, this will cause $E_\Vert$ to drop
until $E_\Vert \approx
E_{av}$~\cite{Rosenbluth:1997,Breizman:2014}. This second root, which
we will refer to as the RE root, is stable for $T_e < T_{av}$, and
leads to the formation of a current plateau. Thus, a sufficient
condition to avoid RE formation is to maintain $T_e \gtrsim
T_{av}$. The primary challenge is to identify a solution whereby $T_e
\gtrsim T_{av}$ while simultaneously adhering to the ITER requirement
of a current quench timescale in the range of 50-150~ms.~\cite{Hollmann-etal-PoP-2015}

The challenge of simultaneously satisfying these two constraints is
made evident in the power balance curves illustrated in
Fig.~\ref{fig:runaway-avoidance}. Here, the Ohmic heating rate is
plotted along with the radiative cooling rate $P_{rad}$ as a function
of the electron temperature $T_e$. The bulk plasma heating can be
estimated by multiplying the parallel electric field sketched in
Fig.~\ref{fig:RE-heating} by the plasma current density. For the Ohmic
root, this leads to the familiar expression $P_\eta \equiv \eta
j^2_\Vert$. For the RE root, this leads to the net energy transferred
to the plasma being given by $P_{RE} = E_{av} j_\parallel$. While a
small fraction of this energy will be lost via radiative losses in the
channels of synchrotron radiation~\cite{Martin:2000, guo2017phase},
bremsstrahlung~\cite{Embreus-etal-NJoP-2016}, and line
emission~\cite{garland-etal-pop-2020,garland-etal-pop-2022}, the
majority of this energy will be collisionally transferred to the bulk
electrons. Thus, at steady state the heating of the bulk electrons
will be bounded from above by $P_{RE} = E_{av} j_\parallel$ when on
the RE root. This estimate of RE heating of the bulk will be a gross overestimate for a recombined plasma, where the REs will primarily collide with bound electrons. The recombined limit, while critical to the treatment of power balance before the flush of REs, will have a negligible impact on the present analysis since we will be interested in a post-flush plasma, where the current is carried by near bulk electrons. For given atomic densities of deuterium and neon
impurities, the collisional-radiative codes
FLYCHK~\cite{chung2005flychk} (for D) and
ATOMIC~\cite{Fontes-etal-JPB-2015} (for Ne) are used to compute the
charge state distribution and the radiative cooling rate, in the
steady-state approximation, as a function of $T_e.$ The free electron
density $n_e$ is then found from quasineutrality. The charge state
distribution is then fed into the avalanche threshold evaluation using
the runaway vortex O-X merger model~\cite{mcdevitt2018relation}, which
accounts for the partial screening effect using the collisional
friction and pitch angle scattering rates given in Hesslow, et
al.~\cite{Hesslow:2017}. This latter step yields an
  estimate of the avalanche threshold as a function of the plasma
  composition. It is interesting to note that at very low $T_e,$
there is a sizable neutral population and the electron-neutral
collisions can contribute significantly to collisional
friction~\cite{frost1961conductivity, schweitzer1966electrical,
  zhdanov2002transport}. This is reflected by the enhanced Ohmic
heating at the low $T_e$ end in Fig.~\ref{fig:runaway-avoidance},
where the Ohmic heating power, after factoring in the electron-neutral
collisions, deviates from the $T_e^{-3/2}$ scaling that is predicted
from the Spitzer resistivity.

Recall that a mitigated post-thermal-quench plasma is radiatively
clamped to low $T_e,$ likely in the range of a few eVs, and the purge
of neon by massive deuterium injection involves a further cooling of
$T_e,$ so the reheating of the bulk plasma necessarily starts from the
very low $T_e$ end, most likely below the first peak of the radiative
cooling rate curve shown in Fig.~\ref{fig:runaway-avoidance}, which is
set by deuterium, not the neon impurity. For high enough deuterium
density $n_D$ and at modest plasma current density,
Ohmic heating may not be able to overcome this first peak in the
radiative cooling curve, and there is no significant reheating of
$T_e$ possible. This is shown by the solid blue curve (radiative cooling)
in Fig.~\ref{fig:runaway-avoidance}(d) in comparison with the
dotted-dash line (Ohmic heating). It is of interest to note that the
deuterium radiative peak, in the case of $n_D=10^{22}$~m$^{-3},$ is
very close to the $P_\eta$ curve in
Fig.~\ref{fig:runaway-avoidance}(a,b). If $j_\parallel$ is dropped
from 2~MA/m$^2$ to 1~MA/m$^2$ in these two cases, $P_\eta$ will also
cross the deuterium radiative cooling peak. For the deuterium
radiation peak to safely stay below $P_\eta$, the deuterium density
$n_D$ must be lower, by an amount that scales with
$j_\parallel^{1/2}.$ For discharges that satisfy this constraint,
the mostly deuterium plasma will be reheated above the deuterium
peak, which is around $T_e=1.2$~eV. This deuterium density constraint
is a necessary, but generally not sufficient condition, for the plasma
to be reheated enough to avoid runaways. The complication comes
from the presence of remnant impurities.

In Fig.~\ref{fig:runaway-avoidance}(a,b), one can see that the
presence of neon impurities, as small as 1-5\% in fractional number
density, introduces a second radiative cooling peak in the range of
$T_e \approx 30\;\text{eV}$. The first crossing point between the
radiative cooling ($P_{rad}$) curve and the Ohmic heating ($P_\eta$)
curve marks the critical electron temperature $T_{reheat}$ that the
reheating of the plasma will be bounded from above. From
Fig.~\ref{fig:runaway-avoidance}(a), we find that with high enough
$n_{neon}$ ($5\%$ for this case), $T_{reheat}$ is in the range of a
few eV to 30 eV. This suggests an in-range Ohmic current quench time,
but avalanche is unavoidable because $T_{reheat} < T_{av}$ for all
three densities.

To further quantify this concept,
we recall that the parallel electric field at
$T_{reheat}$ for an Ohmic plasma (i.e.  the plasma current is purely
Ohmic), is simply $E_{reheat} \equiv E_{\parallel}(\eta) = \eta
j_\parallel$. We can plot the $P_{RE}=E_{av} j_\parallel$ in the same
plot, and the ratio of $P_\eta(T_{reheat}) = \eta(T_{reheat})
j_\parallel^2$ and $P_{RE}$ is just the ratio of $E_{reheat}$ and
$E_{av}.$ Equivalently, we can cast the ratio of $E_{reheat}$ and
$E_{av}$ in terms of the $T_{av}/T_{reheat},$ with $T_{av}$ the
intercept of the runaway heating curve $P_{RE}$ and the Ohmic heating
curve $P_\eta.$ Since $E_\parallel(\eta) = \eta j_\parallel \propto
Z_{eff}/T_e^{3/2}$ with $Z_{eff}$ the effective ion charge of the
plasma, one finds
\begin{align}
  \frac{E_{reheat}}{E_{av}} =
  \frac{Z_{eff}(T_{reheat})}{Z_{eff}(T_{av})} \left(\frac{T_{av}}{T_{reheat}}\right)^{3/2}
  .
\end{align}
Figure~\ref{fig:runaway-avoidance}(b) reveals that even with five times
lower neon densities, $n_{neon}=10^{20}$~m$^{-3}$ (solid blue line)
and $n_{neon}=10^{19}$~m$^{-3}$ (solid red line), which correspond to
fractional number density of 1\% for neon impurities in a deuterium
plasma, $E_{reheat}/E_{av} \ge 10.$ For such a large parallel electric
field, we anticipate robust runaway current reconstitution via the avalanche mechanism.

To safely avoid runaways, $E_\parallel=\eta(T_e) j_\parallel$ should
stay below $E_{av},$ which corresponds to $T_{av} < T_{reheat}.$ From
Fig.~\ref{fig:runaway-avoidance}(a,b), we find that only the case of
lowest $n_D$ ($10^{20}$~m$^{-3}$) and impurity content
($n_{Neon}/n_D=1$\%) satisfies this requirement. And when it does, the
plasma actually recovers from the disruption by reaching electron
temperatures in excess of one keV. This could be a favorable outcome
in a tokamak that offers sufficiently fast positional control to avoid
vertical displacement events (VDEs).  In an ITER-like reactor,
reheating of the plasma with less plasma current simply implies a hot
VDE due to the long wall time of the vacuum vessel, which
could lead to a larger halo current.

If the goal is to terminate the plasma for a shut down of the reactor,
the more desirable scenario lies with much reduced impurity radiation,
but high deuterium density to prevent the plasma from achieving
electron temperatures in excess of a keV.
The limiting case is $n_{neon}=0$ is shown in
Fig.~\ref{fig:runaway-avoidance}(c,d).  One can see that
$n_D=10^{21}$~m$^{-3}$ (solid red curve) is high enough to force
$T_{av} < T_{reheat},$ so the Ohmic electric field stays below the
avalanche threshold. The choice of even higher deuterium densities, for example,
the blue curves in Fig.~\ref{fig:runaway-avoidance}(c) for
$n_D=10^{22}$~m$^{-3},$ offers the intriguing possibility of a lower
$T_{reheat}$ with an Ohmic electric field that is marginally above the
avalanche threshold electric field at $j=2$~MA/m$^2.$ The $T_{reheat}$
is more consistent with a current quench duration of 100~ms envisioned
for ITER, which is in the range of 10-15~eV or so for a deuterium
plasma.  This promising prospect is complicated by the fact that as
the plasma current density drops from 2~MA/m$^2$ to 1~MA/m$^2,$ the
reduction in Ohmic heating power would lead to a radiatively clamped
$T_e$ below the deuterium peak around 1.2~eV in a plasma of
$n_D=10^{22}$~m$^{-3},$ resulting in an Ohmic electric field
significantly above avalanche threshold, see
Fig.~\ref{fig:runaway-avoidance}(d). To avoid the avalanche growth of
runaways during the current quench, one thus relies on (1) the
current-carrying plasma shrinking in size as $I_p$ decays but
maintaining comparable $j_\Vert,$ or (2) a way to dynamically reduce the
plasma particle density as $I_p$ and $j_\Vert$ decay in time.

A number of observations can be made here on both (1) and (2).  For
(1), it is indeed the case that as the toroidal plasma current $I_p$
decreases, the current-carrying plasma column does shrink. The
resulting change in $j_\parallel$ is modest, at most by a factor of
two in an ITER-like plasma initially with 15 MA of plasma current.  In
a goldilocks situation with $T_e$ fixed, a factor of 2 drop in
$j_\parallel$ produces a factor of 4 decrease in $P_\eta.$ Since the
deuterium radiative cooling rate scales with the product of the ion
and electron densities, which is approximately equal in the $T_e\ge
10$~eV range, in order to balance the reduced Ohmic heating rate, the $n_D$
would have to be reduced by a factor of 2 as well.  In practice, the
more likely scenario is that the reduced Ohmic heating due to a lower
$j_\parallel$ would lower $T_e,$ boosting the deuterium radiative
power loss rate in the temperature range of $T_e=10-30$~eV. This would
further aggravate the need to further reduce $n_D.$ Reduction of $n_D$
in the temperature range of $T_e \approx 10-20$~eV can only be achieved
via plasma transport, which can be sustained in a discharge if
particle pumping at the chamber boundary is maintained in a
post-thermal-quench plasma.

The potential remedy to possibly impede a drop of $T_e$ with a decreasing
$j_\parallel$, with edge plasma likely most susceptible to a substantial drop of $T_e$, lies with physical mechanisms that can reduce plasma
cooling with a decreasing $T_e.$ In the targeted range of $T_e
\approx 10-20$~eV, neon radiation intensity rapidly decreases with a
decreasing $T_e.$ This suggests the mitigating role of neon
impurities. By contrasting the radiation intensity of deuterium and
neon around $T_e=20$~eV at fixed $n_e=10^{22}$/m$^3,$ one finds that a fractional number density of
$10^{-5}$ for the neon impurity would have the neon impurity radiative
cooling rate twice that of the bulk deuterium plasma.  Along
the same line, if the neon fractional number density is $10^{-6},$ the
neon radiation would be 1/5 of the deuterium's, and it would have a
negligible offsetting effect in reducing the cooling rate as $T_e$
drops.


\begin{figure}
\begin{centering}
\subfigure[]{\includegraphics[scale=0.25]{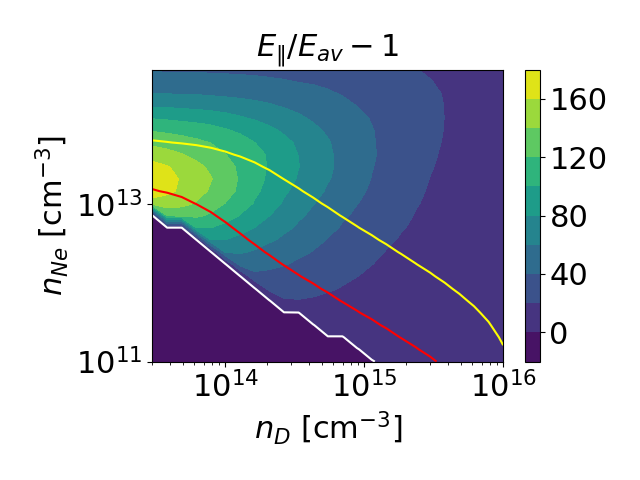}}
\subfigure[]{\includegraphics[scale=0.25]{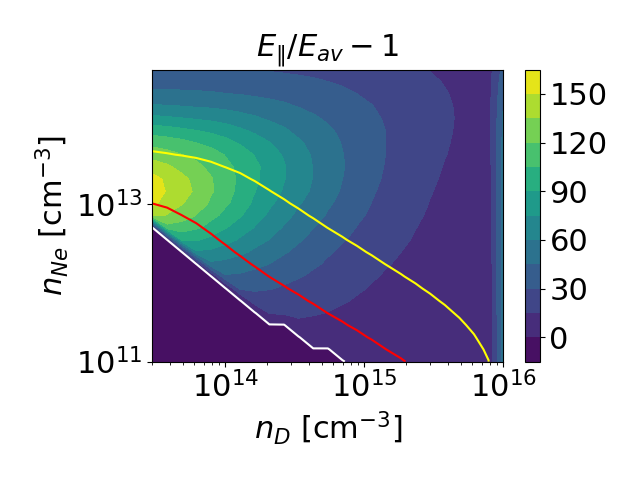}}
\subfigure[]{\includegraphics[scale=0.25]{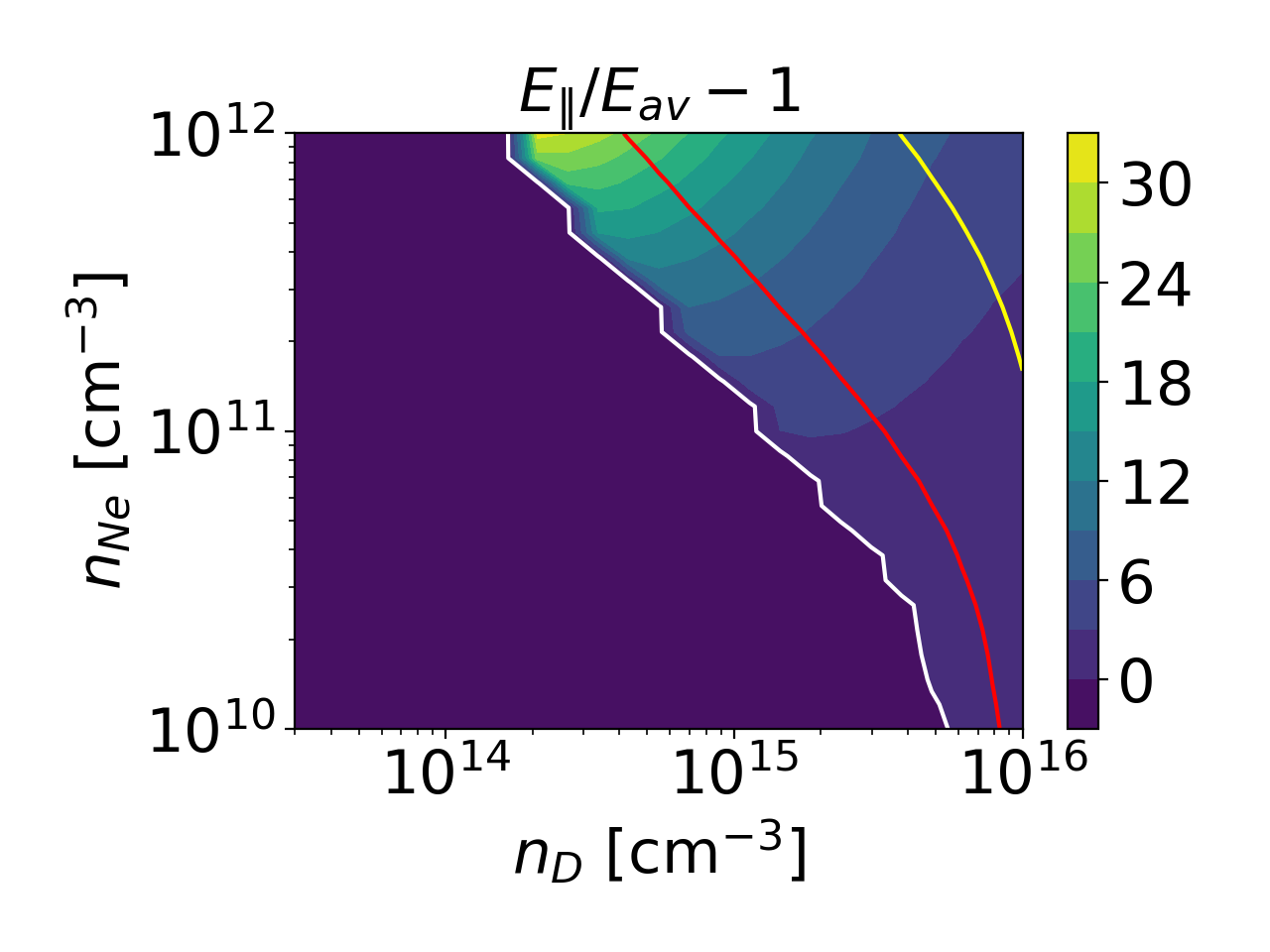}}
\subfigure[]{\includegraphics[scale=0.25]{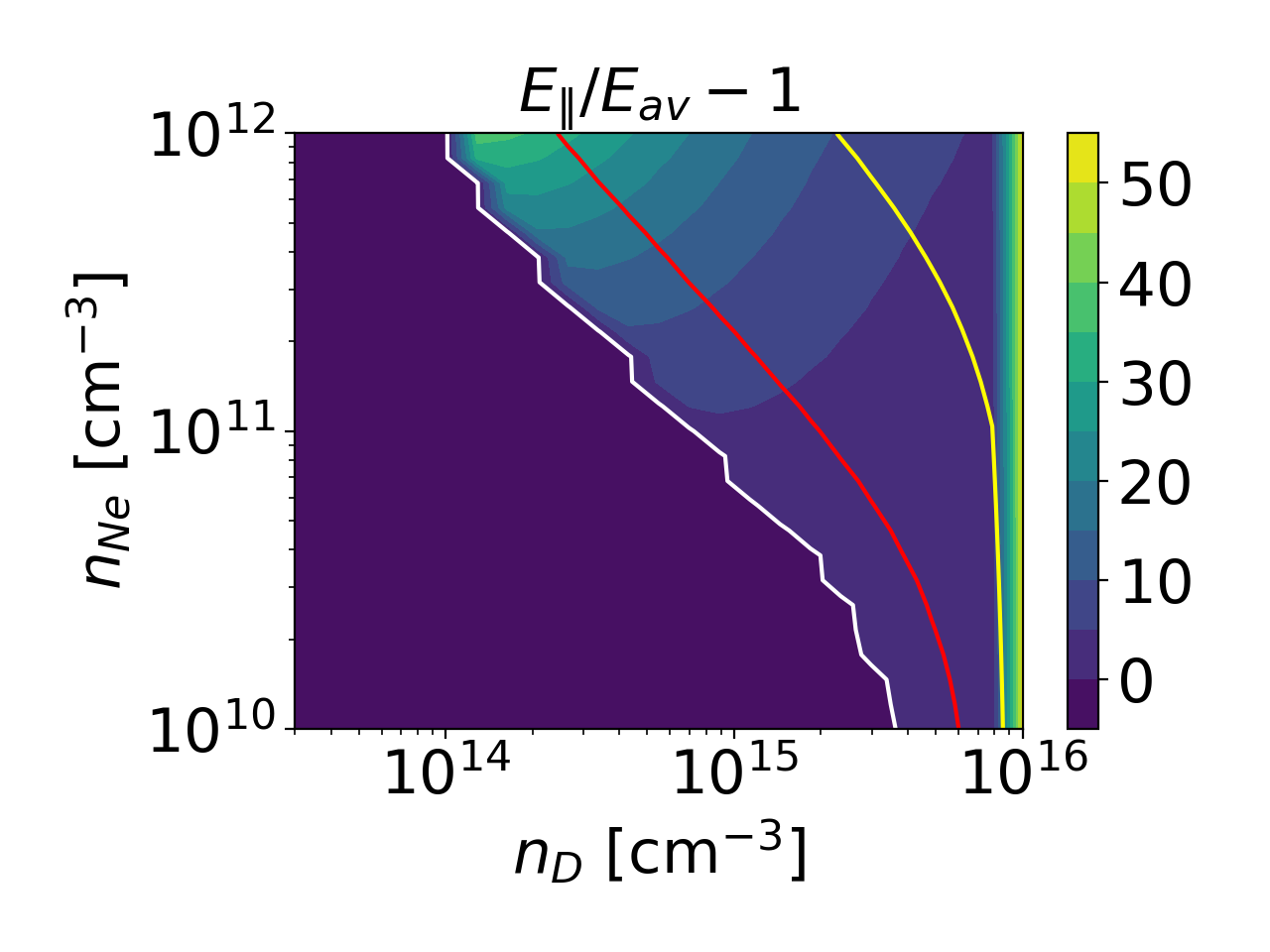}}
\par\end{centering}
\caption{Comparison of the parallel electric field with the avalanche threshold. The yellow contour indicates the location of the $T=10\;\text{eV}$ contour, the red contour indicates the location of the $T=20\;\text{eV}$ contour, and the white contour is for $E_\Vert/E_{av} - 1 = 0$. Panels (a) and (c) are for $2\;\text{MA}/\text{m}^{2}$ and panels (b) and (d) are for $1.5\;\text{MA}/\text{m}^{2}$.}
\label{fig:I5}
\end{figure}

The case studies shown so far clarify the basic physics considerations
and the resulting constraints on the plasma regime for avoiding
runaway avalanche in a post-thermal-quench plasma.  Next we perform a
more comprehensive scan to demarcate the preferred operational regime
in terms of $(n_D, n_{neon}).$ Two derived quantities will be used to
characterize the operational regime. These are ${\cal{E}} \equiv
E_{reheat}/E_{av}-1$ and $T_{reheat},$ all of which were previously
explained in the text and computed in Fig.~\ref{fig:runaway-avoidance}. The result of this calculation is shown in Fig. \ref{fig:I5} for two
different current densities. Two temperature contours are
also plotted, where to remain within the current decay time targeted
by ITER the electron temperature should remain roughly in the range of
$T_e \approx 10-20\;\text{eV}$. 
Considering first a high current density case [see Fig. \ref{fig:I5}(a)] with $j_\parallel = 2\;\text{MA}/\text{m}^{-2}$,
it is evident that the system will remain well above the avalanche
threshold unless a near complete purge of the neon is
present. Furthermore, for low to modest deuterium densities ($n_D
\lesssim 2\times 10^{21}\;\text{m}^{-3}$) the regions below the
avalanche threshold (white contour in Fig. \ref{fig:I5}) coincide with electron temperature in excess of
$100\;\text{eV}$, implying that these cases would have exceptionally long
current decay times. At higher deuterium density a solution near the
avalanche threshold with a temperature in the range of $T_e \approx
10-20\;\text{eV}$ is present though it requires a near complete purge
of the neon. At a modestly lower current density of
$j_\parallel = 1.5\;\text{MA}/\text{m}^{2}$ (see Fig. \ref{fig:I5}(b)), Ohmic heating is unable to overcome the deuterium peak at the highest deuterium density considered. This leads to the plasma recombining, yielding $E_\parallel / E_{av} \gg 1$.
This has the effect of shifting the region with
$E_\parallel/E_{av} \sim 1$ to lower deuterium densities. Hence, the target deuterium density will depend on the local current density of the plasma. Focusing on the very low neon density regime [Figs. \ref{fig:I5}(b) and (d)],
it is evident that even at very low neon densities,
there is no solution below the avalanche threshold with an electron
temperature less than $20\;\text{eV}$. It is, however, apparent that a
solution with the electric field within a factor of two of the
avalanche threshold is present at high deuterium density. Although
this cannot avoid the avalanche growth of runaway electrons, it does
lead to higher poloidal flux consumption in growing the runaway
population, which has the favorable effect of reducing the plasma
current after runaway reconstitution.



In conclusion, the plasma power balance in a post-thermal-quench
plasma 
places a rigorous constraint on the plasma regime in which
runaways can be avoided or minimized. Specifically, unless a current
quench duration of greater than 150~ms can be
tolerated, there does not appear to be a $(n_D, n_{Neon})$ regime in
which runaway avalanche can be completely avoided.  Within the known
ITER constraint for current quench duration
the high $n_D$ but negligibly low $n_{Neon}$ regime can
deliver the desired current quench time while minimizing the runaway
current, by reaching an Ohmic parallel electric field that is above,
but close to, the avalanche threshold electric field.
The accessibility of such a regime poses a pertinent challenge for future experiments. There is the possibility that radiation trapping by a cold/dense edge
  of the high density deuterium plasma~\cite{vallhagen2022effect} can increase the ionization fraction by photo-ionization/excitation and thus
  shift up the $T_e$ for reduced radiative cooling due to higher ionization fraction. This could produce
  a broader `channel' where $E_\Vert/E_{av} \gtrsim 1$ with the current decay rate within the targeted range.
  The exact extent of this effect remains to be studied in the future.


We thank the U.S. Department of Energy Office of Fusion Energy
Sciences and Office of Advanced Scientific Computing Research for
support under the Tokamak Disruption Simulation (TDS) Scientific
Discovery through Advanced Computing (SciDAC) project, and the Base
Theory Program, both at Los Alamos National Laboratory (LANL) under
contract No. 89233218CNA000001. This work was also
  supported by the Research and Development Program of the Korea
  Institute of Fusion Energy (KFE), which is funded by the Ministry of
  Science and ICT of the Republic of Korea (No. KFE-EN2141–8).  Our
computer simulations used resources of the National Energy Research
Scientific Computing Center (NERSC), a U.S. Department of Energy
Office of Science User Facility operated under Contract
No. DE-AC02-05CH11231 and the Los Alamos National Laboratory
Institutional Computing Program, which is supported by the
U.S. Department of Energy National Nuclear Security Administration
under Contract No. 89233218CNA000001.


\begin{thebibliography}{30}
\expandafter\ifx\csname natexlab\endcsname\relax\def\natexlab#1{#1}\fi
\expandafter\ifx\csname bibnamefont\endcsname\relax
  \def\bibnamefont#1{#1}\fi
\expandafter\ifx\csname bibfnamefont\endcsname\relax
  \def\bibfnamefont#1{#1}\fi
\expandafter\ifx\csname citenamefont\endcsname\relax
  \def\citenamefont#1{#1}\fi
\expandafter\ifx\csname url\endcsname\relax
  \def\url#1{\texttt{#1}}\fi
\expandafter\ifx\csname urlprefix\endcsname\relax\def\urlprefix{URL }\fi
\providecommand{\bibinfo}[2]{#2}
\providecommand{\eprint}[2][]{\url{#2}}

\bibitem[{\citenamefont{Hender et~al.}(2007)\citenamefont{Hender, Wesley,
  Bialek, Bondeson, Boozer, Buttery, Garofalo, Goodman, Granetz, Gribov
  et~al.}}]{Hender-etal-nf-2007}
\bibinfo{author}{\bibfnamefont{T.}~\bibnamefont{Hender}},
  \bibinfo{author}{\bibfnamefont{J.}~\bibnamefont{Wesley}},
  \bibinfo{author}{\bibfnamefont{J.}~\bibnamefont{Bialek}},
  \bibinfo{author}{\bibfnamefont{A.}~\bibnamefont{Bondeson}},
  \bibinfo{author}{\bibfnamefont{A.}~\bibnamefont{Boozer}},
  \bibinfo{author}{\bibfnamefont{R.}~\bibnamefont{Buttery}},
  \bibinfo{author}{\bibfnamefont{A.}~\bibnamefont{Garofalo}},
  \bibinfo{author}{\bibfnamefont{T.}~\bibnamefont{Goodman}},
  \bibinfo{author}{\bibfnamefont{R.}~\bibnamefont{Granetz}},
  \bibinfo{author}{\bibfnamefont{Y.}~\bibnamefont{Gribov}},
  \bibnamefont{et~al.}, \bibinfo{journal}{Nuclear Fusion}
  \textbf{\bibinfo{volume}{47}}, \bibinfo{pages}{S128} (\bibinfo{year}{2007}),
  \urlprefix\url{http://stacks.iop.org/0029-5515/47/i=6/a=S03}.

\bibitem[{\citenamefont{SOKOLOV}(1979)}]{Sokolov:1979}
\bibinfo{author}{\bibfnamefont{I.}~\bibnamefont{SOKOLOV}},
  \bibinfo{journal}{JETP Letters} \textbf{\bibinfo{volume}{29}},
  \bibinfo{pages}{218} (\bibinfo{year}{1979}).

\bibitem[{\citenamefont{Jayakumar et~al.}(1993)\citenamefont{Jayakumar,
  Fleischmann, and Zweben}}]{Jayakumar:1993}
\bibinfo{author}{\bibfnamefont{R.}~\bibnamefont{Jayakumar}},
  \bibinfo{author}{\bibfnamefont{H.}~\bibnamefont{Fleischmann}},
  \bibnamefont{and} \bibinfo{author}{\bibfnamefont{S.}~\bibnamefont{Zweben}},
  \bibinfo{journal}{Physics Letters A} \textbf{\bibinfo{volume}{172}},
  \bibinfo{pages}{447} (\bibinfo{year}{1993}).

\bibitem[{\citenamefont{Rosenbluth and Putvinski}(1997)}]{Rosenbluth:1997}
\bibinfo{author}{\bibfnamefont{M.}~\bibnamefont{Rosenbluth}} \bibnamefont{and}
  \bibinfo{author}{\bibfnamefont{S.}~\bibnamefont{Putvinski}},
  \bibinfo{journal}{Nuclear Fusion} \textbf{\bibinfo{volume}{37}},
  \bibinfo{pages}{1355} (\bibinfo{year}{1997}).

\bibitem[{\citenamefont{Mart{\'\i}n-Sol{\'\i}s
  et~al.}(2017)\citenamefont{Mart{\'\i}n-Sol{\'\i}s, Loarte, and
  Lehnen}}]{Martin:2017}
\bibinfo{author}{\bibfnamefont{J.}~\bibnamefont{Mart{\'\i}n-Sol{\'\i}s}},
  \bibinfo{author}{\bibfnamefont{A.}~\bibnamefont{Loarte}}, \bibnamefont{and}
  \bibinfo{author}{\bibfnamefont{M.}~\bibnamefont{Lehnen}},
  \bibinfo{journal}{Nuclear Fusion} \textbf{\bibinfo{volume}{57}},
  \bibinfo{pages}{066025} (\bibinfo{year}{2017}).

\bibitem[{\citenamefont{Hesslow et~al.}(2019)\citenamefont{Hesslow,
  Embr{\'e}us, Vallhagen, and F{\"u}l{\"o}p}}]{hesslow2019influence}
\bibinfo{author}{\bibfnamefont{L.}~\bibnamefont{Hesslow}},
  \bibinfo{author}{\bibfnamefont{O.}~\bibnamefont{Embr{\'e}us}},
  \bibinfo{author}{\bibfnamefont{O.}~\bibnamefont{Vallhagen}},
  \bibnamefont{and}
  \bibinfo{author}{\bibfnamefont{T.}~\bibnamefont{F{\"u}l{\"o}p}},
  \bibinfo{journal}{Nuclear Fusion} \textbf{\bibinfo{volume}{59}},
  \bibinfo{pages}{084004} (\bibinfo{year}{2019}).

\bibitem[{\citenamefont{McDevitt et~al.}(2019)\citenamefont{McDevitt, Guo, and
  Tang}}]{mcdevitt2019avalanche}
\bibinfo{author}{\bibfnamefont{C.~J.} \bibnamefont{McDevitt}},
  \bibinfo{author}{\bibfnamefont{Z.}~\bibnamefont{Guo}}, \bibnamefont{and}
  \bibinfo{author}{\bibfnamefont{X.~Z.} \bibnamefont{Tang}},
  \bibinfo{journal}{Plasma Physics and Controlled Fusion}
  \textbf{\bibinfo{volume}{61}}, \bibinfo{pages}{054008}
  (\bibinfo{year}{2019}).

\bibitem[{\citenamefont{de~Vries et~al.}(2011)\citenamefont{de~Vries, Johnson,
  Alper, Buratti, Hender, Koslowski, Riccardo, and
  Contributors}}]{deVries-etal-nf-2011}
\bibinfo{author}{\bibfnamefont{P.}~\bibnamefont{de~Vries}},
  \bibinfo{author}{\bibfnamefont{M.}~\bibnamefont{Johnson}},
  \bibinfo{author}{\bibfnamefont{B.}~\bibnamefont{Alper}},
  \bibinfo{author}{\bibfnamefont{P.}~\bibnamefont{Buratti}},
  \bibinfo{author}{\bibfnamefont{T.}~\bibnamefont{Hender}},
  \bibinfo{author}{\bibfnamefont{H.}~\bibnamefont{Koslowski}},
  \bibinfo{author}{\bibfnamefont{V.}~\bibnamefont{Riccardo}}, \bibnamefont{and}
  \bibinfo{author}{\bibfnamefont{J.-E.} \bibnamefont{Contributors}},
  \bibinfo{journal}{Nuclear Fusion} \textbf{\bibinfo{volume}{51}},
  \bibinfo{pages}{053018} (\bibinfo{year}{2011}),
  \urlprefix\url{http://stacks.iop.org/0029-5515/51/i=5/a=053018}.

\bibitem[{\citenamefont{Ward and Wesson}(1992)}]{ward1992impurity}
\bibinfo{author}{\bibfnamefont{D.}~\bibnamefont{Ward}} \bibnamefont{and}
  \bibinfo{author}{\bibfnamefont{J.}~\bibnamefont{Wesson}},
  \bibinfo{journal}{Nuclear fusion} \textbf{\bibinfo{volume}{32}},
  \bibinfo{pages}{1117} (\bibinfo{year}{1992}).

\bibitem[{\citenamefont{Breizman et~al.}(2019)\citenamefont{Breizman,
  Aleynikov, Hollmann, and Lehnen}}]{breizman2019physics}
\bibinfo{author}{\bibfnamefont{B.~N.} \bibnamefont{Breizman}},
  \bibinfo{author}{\bibfnamefont{P.}~\bibnamefont{Aleynikov}},
  \bibinfo{author}{\bibfnamefont{E.~M.} \bibnamefont{Hollmann}},
  \bibnamefont{and} \bibinfo{author}{\bibfnamefont{M.}~\bibnamefont{Lehnen}},
  \bibinfo{journal}{Nuclear Fusion} \textbf{\bibinfo{volume}{59}},
  \bibinfo{pages}{083001} (\bibinfo{year}{2019}).

\bibitem[{\citenamefont{Paz-Soldan et~al.}(2021)\citenamefont{Paz-Soldan, Reux,
  Aleynikova, Aleynikov, Bandaru, Beidler, Eidietis, Liu, Liu, Lvovskiy
  et~al.}}]{paz2021novel}
\bibinfo{author}{\bibfnamefont{C.}~\bibnamefont{Paz-Soldan}},
  \bibinfo{author}{\bibfnamefont{C.}~\bibnamefont{Reux}},
  \bibinfo{author}{\bibfnamefont{K.}~\bibnamefont{Aleynikova}},
  \bibinfo{author}{\bibfnamefont{P.}~\bibnamefont{Aleynikov}},
  \bibinfo{author}{\bibfnamefont{V.}~\bibnamefont{Bandaru}},
  \bibinfo{author}{\bibfnamefont{M.}~\bibnamefont{Beidler}},
  \bibinfo{author}{\bibfnamefont{N.}~\bibnamefont{Eidietis}},
  \bibinfo{author}{\bibfnamefont{Y.}~\bibnamefont{Liu}},
  \bibinfo{author}{\bibfnamefont{C.}~\bibnamefont{Liu}},
  \bibinfo{author}{\bibfnamefont{A.}~\bibnamefont{Lvovskiy}},
  \bibnamefont{et~al.}, \bibinfo{journal}{Nuclear Fusion}
  \textbf{\bibinfo{volume}{61}}, \bibinfo{pages}{116058}
  (\bibinfo{year}{2021}).

\bibitem[{\citenamefont{Reux et~al.}(2021)\citenamefont{Reux, Paz-Soldan,
  Aleynikov, Bandaru, Ficker, Silburn, Hoelzl, Jachmich, Eidietis, Lehnen
  et~al.}}]{reux2021demonstration}
\bibinfo{author}{\bibfnamefont{C.}~\bibnamefont{Reux}},
  \bibinfo{author}{\bibfnamefont{C.}~\bibnamefont{Paz-Soldan}},
  \bibinfo{author}{\bibfnamefont{P.}~\bibnamefont{Aleynikov}},
  \bibinfo{author}{\bibfnamefont{V.}~\bibnamefont{Bandaru}},
  \bibinfo{author}{\bibfnamefont{O.}~\bibnamefont{Ficker}},
  \bibinfo{author}{\bibfnamefont{S.}~\bibnamefont{Silburn}},
  \bibinfo{author}{\bibfnamefont{M.}~\bibnamefont{Hoelzl}},
  \bibinfo{author}{\bibfnamefont{S.}~\bibnamefont{Jachmich}},
  \bibinfo{author}{\bibfnamefont{N.}~\bibnamefont{Eidietis}},
  \bibinfo{author}{\bibfnamefont{M.}~\bibnamefont{Lehnen}},
  \bibnamefont{et~al.}, \bibinfo{journal}{Physical Review Letters}
  \textbf{\bibinfo{volume}{126}}, \bibinfo{pages}{175001}
  (\bibinfo{year}{2021}).

\bibitem[{\citenamefont{Bandaru et~al.}(2021)\citenamefont{Bandaru, Hoelzl,
  Reux, Ficker, Silburn, Lehnen, Eidietis, Contributors, Team
  et~al.}}]{bandaru2021magnetohydrodynamic}
\bibinfo{author}{\bibfnamefont{V.}~\bibnamefont{Bandaru}},
  \bibinfo{author}{\bibfnamefont{M.}~\bibnamefont{Hoelzl}},
  \bibinfo{author}{\bibfnamefont{C.}~\bibnamefont{Reux}},
  \bibinfo{author}{\bibfnamefont{O.}~\bibnamefont{Ficker}},
  \bibinfo{author}{\bibfnamefont{S.}~\bibnamefont{Silburn}},
  \bibinfo{author}{\bibfnamefont{M.}~\bibnamefont{Lehnen}},
  \bibinfo{author}{\bibfnamefont{N.}~\bibnamefont{Eidietis}},
  \bibinfo{author}{\bibfnamefont{J.}~\bibnamefont{Contributors}},
  \bibinfo{author}{\bibfnamefont{J.}~\bibnamefont{Team}}, \bibnamefont{et~al.},
  \bibinfo{journal}{Plasma Physics and Controlled Fusion}
  \textbf{\bibinfo{volume}{63}}, \bibinfo{pages}{035024}
  (\bibinfo{year}{2021}).

\bibitem[{\citenamefont{Hollmann et~al.}(2015)\citenamefont{Hollmann,
  Aleynikov, Fülöp, Humphreys, Izzo, Lehnen, Lukash, Papp, Pautasso,
  Saint-Laurent et~al.}}]{Hollmann-etal-PoP-2015}
\bibinfo{author}{\bibfnamefont{E.~M.} \bibnamefont{Hollmann}},
  \bibinfo{author}{\bibfnamefont{P.~B.} \bibnamefont{Aleynikov}},
  \bibinfo{author}{\bibfnamefont{T.}~\bibnamefont{Fülöp}},
  \bibinfo{author}{\bibfnamefont{D.~A.} \bibnamefont{Humphreys}},
  \bibinfo{author}{\bibfnamefont{V.~A.} \bibnamefont{Izzo}},
  \bibinfo{author}{\bibfnamefont{M.}~\bibnamefont{Lehnen}},
  \bibinfo{author}{\bibfnamefont{V.~E.} \bibnamefont{Lukash}},
  \bibinfo{author}{\bibfnamefont{G.}~\bibnamefont{Papp}},
  \bibinfo{author}{\bibfnamefont{G.}~\bibnamefont{Pautasso}},
  \bibinfo{author}{\bibfnamefont{F.}~\bibnamefont{Saint-Laurent}},
  \bibnamefont{et~al.}, \bibinfo{journal}{Physics of Plasmas}
  \textbf{\bibinfo{volume}{22}}, \bibinfo{pages}{021802}
  (\bibinfo{year}{2015}), \eprint{https://doi.org/10.1063/1.4901251},
  \urlprefix\url{https://doi.org/10.1063/1.4901251}.

\bibitem[{\citenamefont{Lehnen et~al.}(2015)\citenamefont{Lehnen, Aleynikova,
  Aleynikov, Campbell, Drewelow, Eidietis, Gasparyan, Granetz, Gribov, Hartmann
  et~al.}}]{lehnen2015disruptions}
\bibinfo{author}{\bibfnamefont{M.}~\bibnamefont{Lehnen}},
  \bibinfo{author}{\bibfnamefont{K.}~\bibnamefont{Aleynikova}},
  \bibinfo{author}{\bibfnamefont{P.}~\bibnamefont{Aleynikov}},
  \bibinfo{author}{\bibfnamefont{D.}~\bibnamefont{Campbell}},
  \bibinfo{author}{\bibfnamefont{P.}~\bibnamefont{Drewelow}},
  \bibinfo{author}{\bibfnamefont{N.}~\bibnamefont{Eidietis}},
  \bibinfo{author}{\bibfnamefont{Y.}~\bibnamefont{Gasparyan}},
  \bibinfo{author}{\bibfnamefont{R.}~\bibnamefont{Granetz}},
  \bibinfo{author}{\bibfnamefont{Y.}~\bibnamefont{Gribov}},
  \bibinfo{author}{\bibfnamefont{N.}~\bibnamefont{Hartmann}},
  \bibnamefont{et~al.}, \bibinfo{journal}{Journal of Nuclear materials}
  \textbf{\bibinfo{volume}{463}}, \bibinfo{pages}{39} (\bibinfo{year}{2015}).

\bibitem[{com()}]{comment:force-free}
\bibinfo{howpublished}{A post-thermal-quench plasma has lost most of its
  thermal energy and hence pressure, so force-free is a good approximation and
  $\mathbf{j}\approx j_\parallel \mathbf{B}/B.$}.

\bibitem[{\citenamefont{Breizman}(2014)}]{Breizman:2014}
\bibinfo{author}{\bibfnamefont{B.~N.} \bibnamefont{Breizman}},
  \bibinfo{journal}{Nuclear Fusion} \textbf{\bibinfo{volume}{54}},
  \bibinfo{pages}{072002} (\bibinfo{year}{2014}).

\bibitem[{\citenamefont{Mart{\i}n-Sol{\i}s
  et~al.}(2000)\citenamefont{Mart{\i}n-Sol{\i}s, S{\'a}nchez, and
  Esposito}}]{Martin:2000}
\bibinfo{author}{\bibfnamefont{J.}~\bibnamefont{Mart{\i}n-Sol{\i}s}},
  \bibinfo{author}{\bibfnamefont{R.}~\bibnamefont{S{\'a}nchez}},
  \bibnamefont{and} \bibinfo{author}{\bibfnamefont{B.}~\bibnamefont{Esposito}},
  \bibinfo{journal}{Physics of Plasmas} \textbf{\bibinfo{volume}{7}},
  \bibinfo{pages}{3814} (\bibinfo{year}{2000}).

\bibitem[{\citenamefont{Guo et~al.}(2017)\citenamefont{Guo, McDevitt, and
  Tang}}]{guo2017phase}
\bibinfo{author}{\bibfnamefont{Z.}~\bibnamefont{Guo}},
  \bibinfo{author}{\bibfnamefont{C.~J.} \bibnamefont{McDevitt}},
  \bibnamefont{and} \bibinfo{author}{\bibfnamefont{X.-Z.} \bibnamefont{Tang}},
  \bibinfo{journal}{Plasma Physics and Controlled Fusion}
  \textbf{\bibinfo{volume}{59}}, \bibinfo{pages}{044003}
  (\bibinfo{year}{2017}).

\bibitem[{\citenamefont{Embr{\'{e}}us et~al.}(2016)\citenamefont{Embr{\'{e}}us,
  Stahl, and Fülöp}}]{Embreus-etal-NJoP-2016}
\bibinfo{author}{\bibfnamefont{O.}~\bibnamefont{Embr{\'{e}}us}},
  \bibinfo{author}{\bibfnamefont{A.}~\bibnamefont{Stahl}}, \bibnamefont{and}
  \bibinfo{author}{\bibfnamefont{T.}~\bibnamefont{Fülöp}},
  \bibinfo{journal}{New Journal of Physics} \textbf{\bibinfo{volume}{18}},
  \bibinfo{pages}{093023} (\bibinfo{year}{2016}),
  \urlprefix\url{https://doi.org/10.1088/1367-2630/18/9/093023}.

\bibitem[{\citenamefont{Garland et~al.}(2020)\citenamefont{Garland, Chung,
  Fontes, Zammit, Colgan, Elder, McDevitt, Wildey, and
  Tang}}]{garland-etal-pop-2020}
\bibinfo{author}{\bibfnamefont{N.~A.} \bibnamefont{Garland}},
  \bibinfo{author}{\bibfnamefont{H.-K.} \bibnamefont{Chung}},
  \bibinfo{author}{\bibfnamefont{C.~J.} \bibnamefont{Fontes}},
  \bibinfo{author}{\bibfnamefont{M.~C.} \bibnamefont{Zammit}},
  \bibinfo{author}{\bibfnamefont{J.}~\bibnamefont{Colgan}},
  \bibinfo{author}{\bibfnamefont{T.}~\bibnamefont{Elder}},
  \bibinfo{author}{\bibfnamefont{C.~J.} \bibnamefont{McDevitt}},
  \bibinfo{author}{\bibfnamefont{T.~M.} \bibnamefont{Wildey}},
  \bibnamefont{and} \bibinfo{author}{\bibfnamefont{X.-Z.} \bibnamefont{Tang}},
  \bibinfo{journal}{Physics of Plasmas} \textbf{\bibinfo{volume}{27}},
  \bibinfo{pages}{040702} (\bibinfo{year}{2020}),
  \eprint{https://doi.org/10.1063/5.0003638},
  \urlprefix\url{https://doi.org/10.1063/5.0003638}.

\bibitem[{\citenamefont{Garland et~al.}(2022)\citenamefont{Garland, Chung,
  Zammit, McDevitt, Colgan, Fontes, and Tang}}]{garland-etal-pop-2022}
\bibinfo{author}{\bibfnamefont{N.~A.} \bibnamefont{Garland}},
  \bibinfo{author}{\bibfnamefont{H.-K.} \bibnamefont{Chung}},
  \bibinfo{author}{\bibfnamefont{M.~C.} \bibnamefont{Zammit}},
  \bibinfo{author}{\bibfnamefont{C.~J.} \bibnamefont{McDevitt}},
  \bibinfo{author}{\bibfnamefont{J.}~\bibnamefont{Colgan}},
  \bibinfo{author}{\bibfnamefont{C.~J.} \bibnamefont{Fontes}},
  \bibnamefont{and} \bibinfo{author}{\bibfnamefont{X.-Z.} \bibnamefont{Tang}},
  \bibinfo{journal}{Physics of Plasmas} \textbf{\bibinfo{volume}{29}},
  \bibinfo{pages}{012504} (\bibinfo{year}{2022}),
  \eprint{https://doi.org/10.1063/5.0071996},
  \urlprefix\url{https://doi.org/10.1063/5.0071996}.

\bibitem[{\citenamefont{Chung et~al.}(2005)\citenamefont{Chung, Chen, Morgan,
  Ralchenko, and Lee}}]{chung2005flychk}
\bibinfo{author}{\bibfnamefont{H.-K.} \bibnamefont{Chung}},
  \bibinfo{author}{\bibfnamefont{M.}~\bibnamefont{Chen}},
  \bibinfo{author}{\bibfnamefont{W.}~\bibnamefont{Morgan}},
  \bibinfo{author}{\bibfnamefont{Y.}~\bibnamefont{Ralchenko}},
  \bibnamefont{and} \bibinfo{author}{\bibfnamefont{R.}~\bibnamefont{Lee}},
  \bibinfo{journal}{High energy density physics} \textbf{\bibinfo{volume}{1}},
  \bibinfo{pages}{3} (\bibinfo{year}{2005}).

\bibitem[{\citenamefont{Fontes et~al.}(2015)\citenamefont{Fontes, Zhang, Jr,
  Clark, Kilcrease, Colgan, Cunningham, Hakel, Magee, and
  Sherrill}}]{Fontes-etal-JPB-2015}
\bibinfo{author}{\bibfnamefont{C.~J.} \bibnamefont{Fontes}},
  \bibinfo{author}{\bibfnamefont{H.~L.} \bibnamefont{Zhang}},
  \bibinfo{author}{\bibfnamefont{J.~A.} \bibnamefont{Jr}},
  \bibinfo{author}{\bibfnamefont{R.~E.~H.} \bibnamefont{Clark}},
  \bibinfo{author}{\bibfnamefont{D.~P.} \bibnamefont{Kilcrease}},
  \bibinfo{author}{\bibfnamefont{J.}~\bibnamefont{Colgan}},
  \bibinfo{author}{\bibfnamefont{R.~T.} \bibnamefont{Cunningham}},
  \bibinfo{author}{\bibfnamefont{P.}~\bibnamefont{Hakel}},
  \bibinfo{author}{\bibfnamefont{N.~H.} \bibnamefont{Magee}}, \bibnamefont{and}
  \bibinfo{author}{\bibfnamefont{M.~E.} \bibnamefont{Sherrill}},
  \bibinfo{journal}{Journal of Physics B: Atomic, Molecular and Optical
  Physics} \textbf{\bibinfo{volume}{48}}, \bibinfo{pages}{144014}
  (\bibinfo{year}{2015}),
  \urlprefix\url{https://doi.org/10.1088/0953-4075/48/14/144014}.

\bibitem[{\citenamefont{McDevitt et~al.}(2018)\citenamefont{McDevitt, Guo, and
  Tang}}]{mcdevitt2018relation}
\bibinfo{author}{\bibfnamefont{C.~J.} \bibnamefont{McDevitt}},
  \bibinfo{author}{\bibfnamefont{Z.}~\bibnamefont{Guo}}, \bibnamefont{and}
  \bibinfo{author}{\bibfnamefont{X.-Z.} \bibnamefont{Tang}},
  \bibinfo{journal}{Plasma Physics and Controlled Fusion}
  \textbf{\bibinfo{volume}{60}}, \bibinfo{pages}{024004}
  (\bibinfo{year}{2018}).

\bibitem[{\citenamefont{Hesslow et~al.}(2017)\citenamefont{Hesslow, Embr\'eus,
  Stahl, DuBois, Papp, Newton, and F\"ul\"op}}]{Hesslow:2017}
\bibinfo{author}{\bibfnamefont{L.}~\bibnamefont{Hesslow}},
  \bibinfo{author}{\bibfnamefont{O.}~\bibnamefont{Embr\'eus}},
  \bibinfo{author}{\bibfnamefont{A.}~\bibnamefont{Stahl}},
  \bibinfo{author}{\bibfnamefont{T.~C.} \bibnamefont{DuBois}},
  \bibinfo{author}{\bibfnamefont{G.}~\bibnamefont{Papp}},
  \bibinfo{author}{\bibfnamefont{S.~L.} \bibnamefont{Newton}},
  \bibnamefont{and}
  \bibinfo{author}{\bibfnamefont{T.}~\bibnamefont{F\"ul\"op}},
  \bibinfo{journal}{Phys. Rev. Lett.} \textbf{\bibinfo{volume}{118}},
  \bibinfo{pages}{255001} (\bibinfo{year}{2017}),
  \urlprefix\url{https://link.aps.org/doi/10.1103/PhysRevLett.118.255001}.

\bibitem[{\citenamefont{Frost}(1961)}]{frost1961conductivity}
\bibinfo{author}{\bibfnamefont{L.}~\bibnamefont{Frost}},
  \bibinfo{journal}{Journal of applied physics} \textbf{\bibinfo{volume}{32}},
  \bibinfo{pages}{2029} (\bibinfo{year}{1961}).

\bibitem[{\citenamefont{Schweitzer and
  Mitchner}(1966)}]{schweitzer1966electrical}
\bibinfo{author}{\bibfnamefont{S.}~\bibnamefont{Schweitzer}} \bibnamefont{and}
  \bibinfo{author}{\bibfnamefont{M.}~\bibnamefont{Mitchner}},
  \bibinfo{journal}{AIAA Journal} \textbf{\bibinfo{volume}{4}},
  \bibinfo{pages}{1012} (\bibinfo{year}{1966}).

\bibitem[{\citenamefont{Zhdanov}(2002)}]{zhdanov2002transport}
\bibinfo{author}{\bibfnamefont{V.}~\bibnamefont{Zhdanov}},
  \bibinfo{journal}{Plasma Physics and Controlled Fusion}
  \textbf{\bibinfo{volume}{44}}, \bibinfo{pages}{2283} (\bibinfo{year}{2002}).

\bibitem[{\citenamefont{Vallhagen et~al.}(2022)\citenamefont{Vallhagen,
  Pusztai, Hoppe, Newton, and F{\"u}l{\"o}p}}]{vallhagen2022effect}
\bibinfo{author}{\bibfnamefont{O.}~\bibnamefont{Vallhagen}},
  \bibinfo{author}{\bibfnamefont{I.}~\bibnamefont{Pusztai}},
  \bibinfo{author}{\bibfnamefont{M.}~\bibnamefont{Hoppe}},
  \bibinfo{author}{\bibfnamefont{S.~L.} \bibnamefont{Newton}},
  \bibnamefont{and}
  \bibinfo{author}{\bibfnamefont{T.}~\bibnamefont{F{\"u}l{\"o}p}},
  \bibinfo{journal}{Nuclear Fusion} \textbf{\bibinfo{volume}{62}},
  \bibinfo{pages}{112004} (\bibinfo{year}{2022}).

\end{thebibliography}

\end{document}